\documentclass[useAMS,usenatbib]{mn2e}
\usepackage{lineno,color}
\RequirePackage{lineno}
\usepackage{amssymb}
\usepackage{amsmath}
\usepackage{txfonts}
\usepackage{graphicx}
\usepackage{natbib}
\usepackage{rotating}
\usepackage{url}
\usepackage{epsfig}
\usepackage{xspace}

\def\ltsima{$\; \buildrel < \over \sim \;$}
\def\simlt{\lower.5ex\hbox{\ltsima}} 
\def\gtsima{$\; \buildrel > \over \sim \;$}
\def\simgt{\lower.5ex\hbox{\gtsima}} 

\def\deg{\hbox{$^\circ$}}
\def\ltsima{$\; \buildrel < \over \sim \;$}
\def\simlt{\lower.5ex\hbox{\ltsima}} 
\def\gtsima{$\; \buildrel > \over \sim \;$}
\def\simgt{\lower.5ex\hbox{\gtsima}} 

\def\deg{\hbox{$^\circ$}}
\def\gr{$\gamma$-ray}
\def\g{$\gamma$}

\def\bi {\begin{itemize}}
\def\ei {\end{itemize}}

\def\xmm{\textit {XMM-Newton}}
\def\szk{\textit {Suzaku}}

\def\swift{\textit {Swift}}
\def\xmmsp{\textit {XMM-Newton }}
\def\psrb{PSR~B1259$-$63}

\def\deg {$^\circ$}

\def\F{{\em Fermi}}

\def\H{H.E.S.S.}
\def\L{LAT}

\def\psrb{PSR~B1259$-$63}

\begin{document}
\title[Multi-wavelength Observations of  \psrb\ around the 2010-2011 Periastron Passage]{Multi-wavelength Observations of  the Binary System \psrb/LS~2883 Around the 2010-2011 Periastron Passage} 
\author[M. Chernyakova et.al.]{M. Chernyakova$^{1,2}$, A.~A. Abdo$^{3}$, A. Neronov $^{4}$, M. V. McSwain$^{5}$, J. Mold\'on$^{6,7}$, M. Rib\'o$^{6}$, \newauthor J.M. Paredes$^{6}$, I. Sushch$^{8,9,10}$, M. de Naurois$^{11}$, U. Schwanke$^{10}$, Y. Uchiyama$^{12,13}$, K. Wood$^{14}$, \newauthor S. Johnston $^{15}$, S. Chaty $^{16,17}$, A. Coleiro $^{16}$
D. Malyshev$^{4,18}$, Iu. Babyk$^{1,2}$\\ 
$^{1}$ Dublin City University, Dublin 9, Ireland\\
$^{2}$ Dublin Institute for advanced studies, 31 Fitzwilliam Place, Dublin 2, Ireland\\
$^{3}$ Operational Evaluation Division,
Institute for Defense Analyses, 4850 Mark Center Drive
Alexandria, VA 22311-1882\\
$^{4}$INTEGRAL Science Data Center, Chemin d'\'Ecogia 16, 
1290 Versoix, Switzerland\\
$^{5}$ Department of Physics, Lehigh University, 16 Memorial Drive East, Bethlehem, PA, USA \\
$^{6}$ Departament d'Astronomia i Meteorologia, Institut de Ci\`ences del
Cosmos, Universitat de Barcelona, IEEC-UB, Mart\'{\i} i Franqu\`es 1,
E-08028 Barcelona, Spain \\
$^{7}$ The Netherlands Institute for Radio Astronomy (ASTRON), 7990-AA Dwingeloo, The Netherlands\\
$^{8}$ Centre for Space Research, North-West University, Potchefstroom Campus, 2520, Potchefstroom, South Africa\\
$^{9}$ Astronomical Observatory of Ivan Franko National University of L'viv, vul. Kyryla i Methodia, 8, L'viv 79005, Ukraine\\
$^{10}$ Institut f\"{u}r Physik, Humboldt-Universit\"{a}t zu Berlin, Newtonstr. 15,  D 12489 Berlin, Germany\\
$^{11}$ Laboratoire Leprince-Ringuet, Ecole Polytechnique, CNRS/IN2P3, F-91128 Palaiseau, France \\ 
$^{12}$ SLAC National Accelerator Laborat 2575 Sand Hill Road, Menlo Park, CA 94025, USA\\  
$^{13}$ Department of Physics, Rikkyo University, Nishi-Ikebukuro 3-34-1,
Toshima-ku, Tokyo, JAPAN  171-8501\\
$^{14}$ U.S. Naval Research Lab 4555 Overlook Ave., SW Washington, DC 20375, USA\\ 
$^{15}$CSIRO Astronomy and Space Science, PO BOX 76, NSW 1710, Australia\\
$^{16}$ Laboratoire AIM (UMR-E 9005 CEA/DSM - CNRS - Universit\'e Paris Diderot), Irfu / Service d'Astrophysique, CEA-Saclay, 91191 Gif-sur-Yvette Cedex, France\\
$^{17}$ Institut Universitaire de France, 103 Boulevard Saint Michel, 75006 Paris, France\\
$^{18}$ Bogolyubov Institute for Theoretical Physics,14-b Metrolohichna street, Kiev, 03680, Ukraine}

\date{Received $<$date$>$  ; in original form  $<$date$>$ }
\pagerange{\pageref{firstpage}--\pageref{lastpage}} \pubyear{2013}

\maketitle
\label{firstpage}
\begin{abstract}
{ We report on broad multi-wavelength observations of the 2010-2011 periastron passage of the \g-ray loud binary system \psrb.  High resolution interferometric radio observations establish extended radio emission trailing the position of the pulsar. Observations with the \F\ \textit{Gamma-ray Space Telescope} reveal GeV \g-ray flaring activity of the system, reaching the spin-down luminosity of the pulsar, around 30 days after periastron. There are no clear signatures of variability at radio, X-ray and TeV energies at the time of the GeV flare. Variability around periastron in the H$\alpha$
emission line, can be interpreted as the gravitational interaction between
the pulsar and the circumstellar disk. The equivalent width of the H$\alpha$ grows
from a few days before periastron until a few days later, and decreases again
between 18 and 46 days after periastron. In near infrared we observe the similar decrease of the equivalent width of Br$\gamma$ line between the 40th and 117th day after the periastron.  For the idealized disk, the variability of the H$\alpha$ line represents the variability of the mass and size of the disk. We discuss
 possible physical relations between the state of the disk and GeV emission under assumption that  GeV flare is directly related to the decrease of the disk size. 
}
\end{abstract}

\section{Introduction} \label{section-intro}

The binary system \psrb\ is comprised of a 47.76 ms radio pulsar in a highly
eccentric orbit ($e \approx 0.87, P \approx 3.4$ years) around the massive main
sequence O9.5Ve star LS~2883 \citep{Johnston1992,distance}. The companion shows
evidence for an equatorial disk in its optical spectrum, and it has generally
been classified as a Be star. The minimum approach between the pulsar and
massive star is about $\sim 0.9$ astronomical unit (AU) \citep{distance}, which
is roughly the size of the equatorial disk \citep{Johnston1992}. The orbital
plane of the pulsar is thought to be inclined with respect to this disk, so it 
crosses the disk plane twice each orbit, just before and just after the
periastron passage \citep[e.g.][]{melatos1995}. A shock interaction between the
relativistic pulsar wind and the wind and photon field of the Be star is
believed to give rise to the observed unpulsed X-ray emission observed
throughout the orbit \citep{Hirayama1999, Chernyakova2006} and the unpulsed
radio and TeV \g-rays observed within a few months of periastron passage
{\citep{Johnston1999,Johnston2005,kirk99,Aharonian_B1259_2004_pass,psrb1259_hess09}.}

The emission from the system varies dramatically as the pulsar moves through
very different environments, making it an excellent test bed for models of a
shocked pulsar wind.  When the pulsar is far from periastron, the highly
linearly polarized pulsed radio emission is detected
\citep{Johnston1999,Johnston2005}. As the pulsar approaches the companion star
( $\sim t_{\rm p} -100$~days, where  $t_{\rm p}$ is the time of periastron passage),
depolarization of the pulsed emission occurs, and the dispersion measure (DM)
and absolute value of the rotation measure (RM) begin to increase while the
pulsed flux density decreases \citep{Johnston1996}. In the range $\sim t_{\rm
p} \pm 15$ days, no pulsed flux is detected \citep{Johnston1996}. The eclipse
of the pulsar during this period is likely due to absorption and severe pulse
scattering by the Be star's disk \citep{Johnston1996}. This eclipse of the
pulsed emission is accompanied by an increase in the unpulsed radio flux
beginning at $\sim t_{\rm p} -30$~days and reaching a maximum at about $t_{\rm p} - 10$~days. It then decreases around the periastron passage before climbing to a second peak about 20 days after the periastron passage
\citep{Johnston1999,Johnston2005}, finally declining over $\sim t_{\rm p} +
100$~days. Unpulsed flux at the two peaks before and after periastron passage can be several
times higher than the value during the periastron passage.

The radio emission has been imaged at scales of AU with the Australian Long
Baseline Array, revealing that it extends up to projected distances of more
than 100 AU close to the periastron passage \citep{Moldon11}. The peak of this
radio emission is detected at projected distances of several tens of AU outside
the binary system (provided unmodeled ionospheric uncertainties are not very
large). This has been the first observational evidence that non-accreting
pulsars orbiting massive stars can produce variable extended radio emission at
AU scales. It must be noted that similar structures have been detected in the
\g-ray binaries LS~5039 and LS~I~+61~303, reinforcing the links between
these three sources and supporting the presence of pulsars in these systems as
well (\citealt{moldon12} and references therein).

In the X-ray band, variable unpulsed emission is observed throughout the orbit
(\citealt{Chernyakova2009} and references therein), even at apastron. For most
of the orbit, the typical X-ray flux is on the order of $F_{\rm X} \sim
10^{-12} \mbox{ erg } \mbox{cm}^{-2} \mbox{ s}^{-1}$. Commencing 20--30 days
prior to periastron passage, there is a sharp rise, reaching 10--20 times the
apastron flux. After that, the  flux decreases by a factor of a few to a local
minimum near or after periastron passage, then to a post-periastron second
maximum similar to the pre-periastron maximum, following which it declines over
100--150 days.  This twofold X-ray maximum broadly resembles the unpulsed radio
light curve.  Moreover, the X-ray pattern is essentially repeated during each
orbit, without large orbit-to-orbit variations. 

Apart from the variable X-ray emission from the binary system itself,
\citet{pavlov} have recently reported the discovery of extended X-ray emission
on distance scales much larger than the size of the binary orbit ($4''$,
corresponding to a projected distance of $\sim 10^{17}$~cm). This indicates
that the pulsar is also powering a ``regular'' pulsar wind nebula during the
periods outside the periastron passage. 

At energies around 1~GeV, the \textit{Energetic Gamma-Ray Experiment Telescope (EGRET)}
produced only an upper limit for the 1994 periastron passage ($F_{\gamma} \leq
9.4 \times 10^{-8}\mbox{ photons } \mbox{ cm}^{-2} \mbox{ s}^{-1}$ for $E\geq$
300~MeV, 95\% confidence \citep{Tavani_B1259_1996}), and there was no
opportunity to observe at these energies during the next four periastron
passages.  However, in TeV \g-rays the system was detected during the 2004 and
2007 periastron passages \citep{Aharonian_B1259_2004_pass, psrb1259_hess09}, 
and flux variations on daily timescales from zero to $10^{-11}$ cm$^{-2}$
s$^{-1}$ were seen for energies $>$0.38~TeV in 2004
\citep{Aharonian_B1259_2004_pass}. A combined lightcurve of those two
periastron passages reveals a hint of two asymmetrical peaks centered around
periastron with a decrease of the flux at the periastron itself.
Peaks of the TeV emission roughly coincide with the flux enhancement observed
in other wavebands as well as the eclipse of the pulsed radio emission. In
2007, the photon flux became notable at the level of 3 standard deviations ($\sigma$) from $\sim t_{\rm
p} - 75$ days  onwards. No TeV detection has been reported far from periastron
\citep{psrb1259_hess09}.

The most recent periastron passage took place on 2010 December 14, 16:40:50.6 (MJD
55544.7). For the first time, we had a chance to observe it in GeV range with
the Large Area Telescope (LAT) of \F\ satellite. No detection at the level of 5$\sigma$ was observed from the
source on daily and weekly   timescales prior to 28 days before the periastron,
$t_{\rm p}-28$~days. Integrating from $t_{\rm p}-28$~days (the typical start of enhanced
X-ray and unpulsed radio flux) to  periastron yielded a clear detection of
excess $\gamma$-ray flux from the source at a 5$\sigma$ level in the energy
range 0.1--1~GeV. The source disappeared after $t_{\rm p}+18$~days and suddenly
become bright again thirty days after the periastron, reaching a flux $\sim$10
times higher than the integrated flux measured during the periastron passage,
\citep{Tam2011,Abdo2011_b1259}. This flare lasted for about 7 weeks with no
corresponding rise in X-ray or radio domains \citep{Abdo2011_b1259}.

Here we report on the multi-wavelength observations of PSR~B1259$-$63 over
$\sim20$ decades of energy, from radio to TeV \gr s, during its 2010 periastron
passage. The paper is organized in the following way: in Sect.~\ref{radio} we
describe radio observations, in Sect.~\ref{infrared} we report NIR observations, in Sect.~\ref{optical} we discuss optical
observations, Sects.~\ref{x-rays}, \ref{fermi}, and \ref{hess} are devoted to
X-ray, GeV and TeV observations, respectively. The discussion of all the
results is in  Sect.~\ref{discussion}, and finally the conclusions are in
Sect.\ref{conclusions}.\\


\section{Radio Observations and Results} \label{radio}

\subsection{Pulse Monitoring with Parkes}

\begin{figure*}
\resizebox{0.9\hsize}{!}{\includegraphics[angle=0]{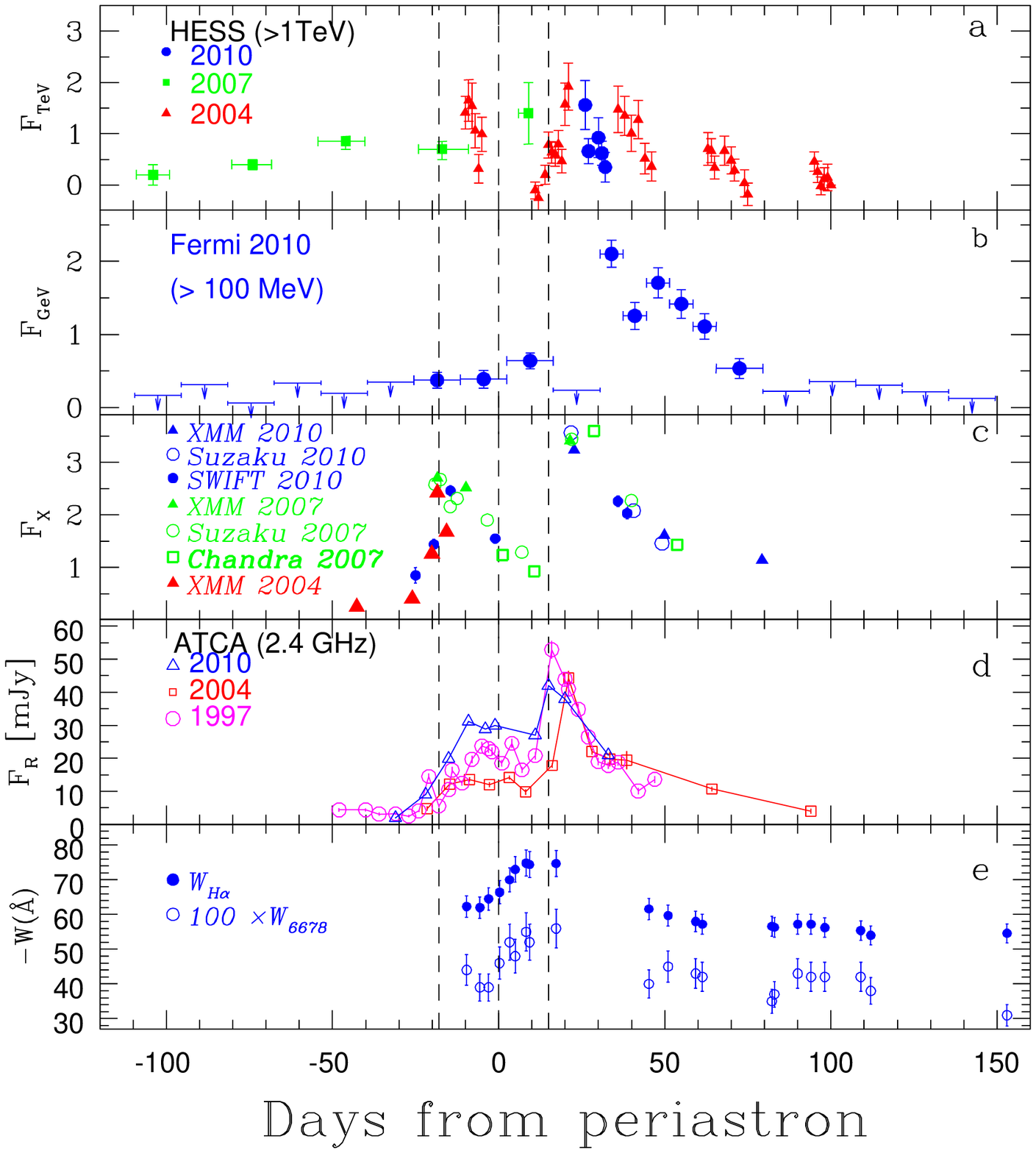}}
\caption{Orbital light curves of \psrb\ around periastron for several
  passages. \textit{ Panel a}: observations by H.E.S.S. in the $E>1$ TeV
  energy range for the 2004, 2007, and 2010 periastron passages
  \citep{Aharonian_B1259_2004_pass, psrb1259_hess09,
    psrb1259_hess13}. Flux is given in $10^{-12}$~{cm}$^{-2}$~s$^{-1}$. \textit{Panel b}: \F-\L\ flux measurements in
  the $E > 100$~MeV energy range for the 2010 periastron passage. Flux is given in $10^{-6}$~cm$^{-2}$~s$^{-1}$. \textit{Panel c}:
  X-ray fluxes from three periastron passages \citep{Abdo2011_b1259,
    Chernyakova2009}. Flux is given in $10^{-11}$~erg~cm$^{-2}$~s$^{-1}$. The typical error of the X-ray data is smaller than the size of the symbols.  \textit{Panel d}: Radio (2.4~GHz) flux densities
  measured at ATCA for the 2010, 2004 and 1997 periastron passages
  \citep{Abdo2011_b1259, Johnston2005,Johnston1999}. Dashed lines correspond to the periastron and to the moments of disappearence (last detection) and reappearence (first detection) of the pulsed emission. \textit{Panel e}: Evolution of the equivalent widths of H$\alpha$ (filled circles) and He I $\lambda6678$ (open circles). $W_{6678}$ is shown multipled by a factor of 100 for easier comparison to $W_{\rm H\alpha}$.}
\label{fig:MW_LC} 
\end{figure*}

\psrb\ is one of the pulsars regularly timed with the Parkes radio telescope as
part of the \F\ timing consortium \citep{DAS08, Weltevrede2010}. In order to
look for changes in the DM and RM and to determine the duration of the eclipse
of the pulsed signal, we monitored the pulsar with Parkes during the 2010
periastron passage. The last detection of  pulsed emission in 2010 happened
eighteen days before the periastron ($t=t_{\rm p}-18$~days). The first observation
with no pulsations happened at  $t_{\rm p}-16$~days. Significant changes in the DM
were observed during  $\sim2$ weeks leading to the disappearance of the pulse.
Pulsation was re-detected on $t_{\rm p}+15$~days at 3.1~GHz and subsequently at
lower frequencies ($t_{\rm p}+16$~days at 1.4~GHz). The moments when the pulsed
emission disappears (last detection) and reappears (first detection) are marked
with dashed lines in Fig.~\ref{fig:MW_LC}.  

\subsection{ATCA Observations}

Monitoring of the unpulsed radio signal was performed with the Australia
Telescope Compact Array (ATCA).  We monitored \psrb\ at frequencies between 1.1
and 10~GHz. A total of twelve observations were collected in the period between
$t_{\rm p}-31 \mbox{days and }t_{\rm p}+55$~days. Unpulsed radio emission was detected
throughout the periastron passage with a behavior similar to that seen in
previous periastron passages \citep{Johnston2005, Abdo2011_b1259}. Panel (d) of
Fig.~\ref{fig:MW_LC} shows the radio light curve for the 1997, 2004 and 2010
periastron passages.

\subsection{LBA observations of PSR~B1259$-$63} \label{sec:VLBI}

\subsubsection{LBA Observations and Data Reduction}

Very Long Baseline Interferometer (VLBI) observations of \psrb\ were conducted
with the Australian Long Baseline Array (LBA) at 2.3~GHz (13~cm) on 2011 January
13 (MJD~55574), from 12:00 to 23:55~UTC. The orbital phase of the binary system
during the observation was 0.243, computed using the ephemerides in
\cite{Wang2004}, and corresponds to 29.3~days after the periastron passage. Six
antennas participated in the observations: ATCA (as a phased array), Ceduna,
Hobart, Mopra, Parkes, and Tidbinbilla (we note that the data from Tidbinbilla
could not be properly calibrated and were not included in the final data
reduction).

The data were obtained with dual circular polarization. ATCA, Mopra, and Parkes
recorded eight 16~MHz subbands (four for each right- and left-handed
polarization) for a total data bit-rate of 512~Mbps per station, and the rest of
the antennas recorded four 16~MHz subbands, for a total bit-rate of 256~Mbps.
The correlation of the data was conducted at the correlator in the International
Centre for Radio Astronomy Research (ICRAR), which produced the final
visibilities with an integration time of 2 seconds. Two passes of the correlator
were conducted: one with all the data and for all sources, and a second one
correlated only during the on-pulse of \psrb\ using pulsar binning. The pulsar
ephemerides were obtained from long-term timing observations conducted with
Parkes.

The observations were performed using phase referencing on the calibrator
J1256$-$6449, located at 1.2$^{\circ}$ from \psrb. J1256$-$6449 was correlated
at $\alpha_{\rm J2000.0}=12^{\rm h} 56^{\rm m} 03\fs4032$ and $\delta_{\rm
J2000.0}=-64\degr 49\arcmin 14\farcs814$, which has an absolute uncertainty in
the ICRF (International Celestial Reference Frame) of 2.8~mas in right ascension
and in declination. The cycle time was 6~minutes, spending half of the time on
the phase calibrator and the target source alternatively. The sources
J1332$-$6646 and J1352$-$4412 were observed as fringe finders.

The data reduction was performed using the NRAO Astronomical Image Processing
System  (AIPS)\footnote{AIPS is available online at http://www.aips.nrao.edu/}.
We applied a priori flagging on telescope off-source times because of antenna
slewing, and ionospheric corrections obtained from total electron content (TEC)
models based on GPS data obtained from the Crustal Dynamics Data Information
System (CDDIS)  data archive\footnote{CDDIS is available online at
http://cddis.nasa.gov/}. The amplitude calibration was performed using the
system temperatures measured at each station, although no antenna gain curves
were available. The amplitude calibration was fixed by scaling the individual
antenna gains by a factor obtained from the phase calibrator and fringe-finder
models. This correction scales the visibility amplitudes between antennas, and
therefore the measured source flux density is not reliable. The phase
calibration was obtained from the phase reference calibrator J1256$-$6449 using
the AIPS task FRING. The phase solutions were applied to the data of \psrb . We
produced an image of \psrb\ using the task IMAGR, with robust parameter 0, and
$uv$ tapering of 8~M$\lambda$ to reduce the weight of the longest and more
noisy baselines. We further calibrated the data with several iteration steps
of  phase only and phase+amplitude self-calibrations. Due to the limited number
of antennas, for amplitude self-calibration we always used integration time
intervals of at least 6 hours (half of the total 12-hour observation time). The
integration time of the phase self-calibration steps was 10 minutes, and this
was reduced to 2 and 1 minutes in the last steps.

The amplitude and phase calibration tables from the phase reference source were
also applied to the on-pulse data of \psrb. In this data set the flux density
of the pulsed emission is slightly enhanced, whereas for the unpulsed emission,
i.e. extended nebula, it remains equal. We produced self-calibrated images
following the same procedure described above. We subtracted the image obtained
using the whole data from the image obtained using the on-pulse data only,
taking into account a scaling factor for the amplitudes. The resulting image
shows a point-like source at the position of the enhanced pulsed emission and
therefore marks the position of the pulsar.

\subsubsection{LBA Results}

The phase-referenced data provide an accurate determination of the position of \psrb\ with respect to the phase calibrator. The measured position of the peak of the source in the ICRF is $\alpha_{\rm J2000.0}=13^{\rm h} 02^{\rm m} 47\fs64239\pm0.3$~mas ($\pm$2.8~mas), and $\delta_{\rm J2000.0}=-63\degr 50\arcmin
08\farcs6267\pm0.3$~mas ($\pm$2.8~mas), where the first set of uncertainties correspond to the formal errors of a Gaussian fit obtained with JMFIT within AIPS, and the uncertainties in parenthesis correspond to the absolute uncertainty of the phase calibrator position in the ICRF. Additional systematic errors due to the unmodeled ionosphere of 1--5~mas are expected. We note that the source is intrinsically extended (see \citealt{Moldon11}) and therefore the astrometry depends on the resolution of the observation, which for the phase-referenced image is $18.4\times14.6$~mas$^2$ at $87$\deg.

\begin{figure}
\center
\resizebox{1.0\hsize}{!}{\includegraphics[angle=0]{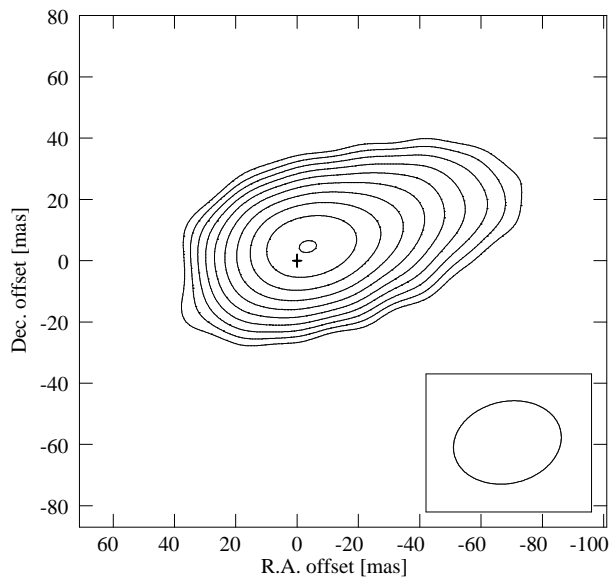}}
\caption{Self-calibrated LBA image of PSR~B1259$-$63 obtained on 2011 January 13
(29 days after periastron passage) at 2.3~GHz frequency. The contours start at 3
times the rms of the image and increase by factors of $2^{1/2}$, up to a peak
signal-to-noise ratio of 68. The (0,0) coordinates correspond to the position of
the pulsar. The cross represents the pulsar position uncertainty at the
5-$\sigma$ level. The synthesized beam, plotted in the lower   right corner, has
a size of $35.5\times26.7$~mas$^2$ and is oriented with P.A. of $-$77.1\deg.}
\label{fig:lba}
\end{figure}

In Fig.~\ref{fig:lba} we show the final self-calibrated image of \psrb\
obtained 
29~days after the periastron passage. The
structure is similar to the one reported in \cite{Moldon11} 21 days after the
previous periastron passage. The source shows a main core and extended diffuse
emission towards the North-West. In this image, for the first time, we have an
accurate position of the pulsar within the unpulsed extended nebula.

The extended emission has a total size of $\sim$50~mas with a position angle
(P.A.) of approximately $-75^{\circ}$. Visual inspection of higher-resolution
images shows that the structure is dominated by a bright compact core and a
diffuse component. We fitted two components to the interferometric $uv$-plane
using the task UVFIT. We found that the core is well fitted by a point-like
component, whereas the diffuse emission is described by a circular Gaussian
component with a FWHM of 20~mas located at $-32.0\pm0.2$ and $8.7\pm0.2$~mas
from the core in right ascension and declination, respectively. It was not
possible to fit an additional component for the pulsar, which is too faint to be
distinguished from the core component with the current noise level.

The position of the pulsar, obtained from the difference between the gated and
the ungated data, is $3.1\pm0.3$~mas in right ascension and $-4.6\pm0.4$~mas in
declination from the core component as seen with the current resolution, for a
total separation of $5.6\pm0.4$~mas. The position of the pulsar is marked with a
cross in Fig.~\ref{fig:lba}, and its size is the estimated uncertainty at
5-$\sigma$ level. The uncertainty in the relative astrometry of the pulsar comes
from the limitation of efficiently subtracting the nebula contribution in the
two images (gated and ungated), mainly because of small differences in the
self-calibration and the determination of the scaling factor for the amplitudes,
which is 0.25. We explored a range of the scaling factors between 0.15 and 0.35
and considered only those images showing a point-like source, i.e. for which the
nebula was completely subtracted. We measured the pulsar position in each image,
which followed a linear displacement as a function of the scaling factor. The
maximum separation of these peak positions was divided by two to estimate the
position uncertainty. An independent estimate of the position uncertainty comes
from the fitting procedure of the JMFIT task within AIPS. We have considered the
higher of these two values in each coordinate as the estimate of the 1-$\sigma$
uncertainty in position, which turns out to be 0.3~mas in right ascension and
0.4~mas in declination. The cross plotted in Fig.~\ref{fig:lba} represents five
times these values to be on the conservative side, given possible differences
related to the self-calibration processes.

\section{Infrared spectroscopy} \label{infrared}

We performed European Southern Observatory (ESO) Very Large Telescope (VLT) observations of PSR\,B1259-63 on Unit Telescope 1 (UT1-Antu), using the Target of Opportunity (ToO) programme ID 086.D-0511 (PI Chaty) dedicated to optical/infrared study of high energy transients in support to {\it Fermi} observations. We obtained low-resolution ($R = \sim 750$) infrared spectra with the Infrared Spectrometer And Array Camera (ISAAC) instrument in near-infrared (NIR, SWS-LR JHKs 1-2 microns) on January 22/23 and March 12, 2011. The ToO was activated in two dates spaced by $\sim 1.5$\,month, in January and March 2011, to compare two different sets of data, and potentially detect spectral variability of the source. Because of the brightness of the NIR counterpart (Ks=7.248), and to avoid non-linearity, the exposure time of individual ISAAC SW spectra was set to 3.55\,s. Observations were carried out under clear conditions with a seeing ranging from $0.35$ to $0.50$\,arcsec. Spectra analysis was standard, including zero correction, flat field correction, extraction, and wavelength calibration, using standard slit spectroscopy routines in the  Image Reduction and Analysis Facility (IRAF) suite \footnote{IRAF is distributed by the National Optical Astronomy
Observatory, which is operated by the Association of Universities for Research
in Astronomy (AURA) under cooperative agreement with the National Science
Foundation. IRAF is available online at http://iraf.noao.edu/ .}.

\begin{figure}
        \center{\includegraphics[width=\columnwidth]
        {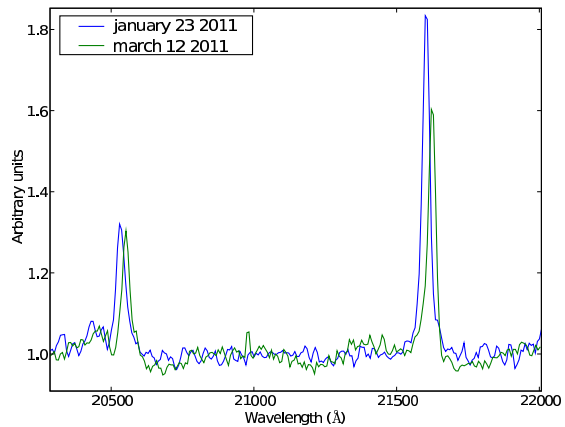}}
        \caption{\label{figure:Kspectra} ESO/VLT NIR Ks band spectra obtained in January 23 and March 12, 2011.}
\end{figure}

\begin{figure}
        \center{\includegraphics[width=\columnwidth]
        {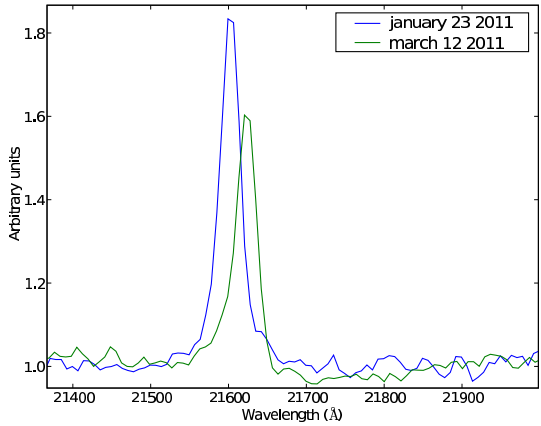}}
        \caption{\label{figure:Brgamma} ESO/VLT NIR Ks band spectra obtained in January 23 and March 12, 2011, enlarged on the region including the Br$\gamma$ line.}
\end{figure}

We show in Figure \ref{figure:Kspectra} both NIR Ks band spectra obtained in January and March 2011, where we clearly see two strong lines in emission: HeI and Br$\gamma$. We show in Figure \ref{figure:Brgamma} a zoom on the region including the Br$\gamma$ line, to see the evolution of its profile between the two observing dates. We also report in Table \ref{table:IRlines} the characteristics of HeI and Br$\gamma$ lines: wavelength, equivalent width (EW) and Full Width Half Maximum (FWHM). We estimate the uncertainty of these parameters to $\sim 10\%$ due to spectra analysis and localisation of the continuum by eye inspection (using the IRAF task {\it splot}). The overall uncertainty on wavelength calibration is due both to the instrumental resolution and to extraction of lamp spectrum, which are of the order of $\sim 0.3$\,nm and $\sim 2$\,nm respectively. Some lines appear at fit wavelength offset with respect to their laboratory position, with an offset greater in January than in March (see Table \ref{table:IRlines}). While this offset might be created by Doppler effect of ejecta outflowing from the decretion disk, since we do not detect any similar offset at optical wavelengths, we can not exclude the possibility that this offset is due to larger uncertainty on wavelength calibration.

By comparing these NIR Ks-band spectra to the optical spectrum exhibiting a strong H$\alpha$ line, and to the series of optical spectra centered on the HeI\,6678 line, we observe the same general tendency both in optical and infrared, i.e. in both cases the EW of H$\alpha$ and Br$\gamma$ lines gradually decrease between January and March 2011 (compare Figure \ref{fig:MW_LC} panel e with Figure \ref{figure:Brgamma}, and Table \ref{tab:optical} with Table \ref{table:IRlines}). In addition, the Br$\gamma$ line is asymmetric (as the HeI\,6678 line), with the blue wing slightly wider in March than in January, which is reminiscent of the asymetric double peaked HeI\,6678 line. This asymmetry may indicate either the presence of a spiral density wave in the disk as suggested by the observations of HeI\,6678 line, or even a truncation of the disk size between the compact object and the companion star (such as described in \cite{okazaki01,Okazaki2011}). We note that the asymmetric wing could also be explained by the presence of ejecta outflowing from the decretion disk.

Finally, we notice that, while the intensity of the HeI line remains stable, the one of the Br$\gamma$ line decreases between January and March observations, which might indicate a change in ionization in the part of the disk where these lines are created.

\begin{table}
\caption{Identified lines in ESO/VLT NIR Ks band spectra obtained in January 23 and March 12, 2011} 
\centering 
\begin{tabular}{c c c c c c} 
\hline\hline 
Date & Line& $\lambda_{lab} (\AA)$ & $\lambda_{fit} (\AA)$ & EW (\AA) & FWHM (\AA)\\ [0.5ex] 
\hline 
January & HeI       & 20580 & 20531 & -10.6 & 35.3 \\
March   & HeI       & 20580 & 20549 & -10.1 & 32.2 \\
January & Br$\gamma$& 21661 & 21601 & -25.2 & 30.3 \\
March   & Br$\gamma$& 21661 & 21622 & -20.4 & 32.0 \\ [1ex]
\hline
\end{tabular}
\label{table:IRlines}
\end{table}

\section{Optical Spectroscopy} \label{optical}

\begin{figure}
\center
\resizebox{0.95\hsize}{!}{\hspace{10mm}\includegraphics[angle=90]{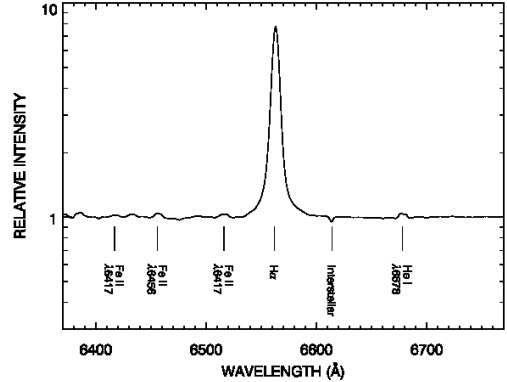}}
\caption{The mean H$\alpha$ spectrum of LS~2883 is plotted with several other faint emission lines.  The relative intensity is plotted on a log scale to emphasize the weaker emission lines. }
\label{fig:meanspec} 
\end{figure}

Spectroscopic observations of LS~2883, the optical counterpart of \psrb, were
performed with the CTIO 1.5m telescope, operated by SMARTS consortium \footnote{http://www.astro.yale.edu/smarts}, between UT dates
2010 December 5 and 2011 May 17.  We used the RC spectrograph in service
observing mode with the standard SMARTS grating setup 47/Ib (grating 47 in
$1^{\rm st}$ order) with the GG495 order sorting filter to achieve a wavelength
coverage of 5630--6940 \AA\ and a resolving power $R = \lambda/\Delta \lambda =
2500$ in the vicinity of the H$\alpha$ line.  We observed LS~2883 for 3
$\times$ 300 seconds each night for a total of 21 nights.  Neon comparison lamp
spectra were obtained before and after the sequence for wavelength
calibration.  The spectra were zero corrected, flat fielded, extracted, and
wavelength calibrated using standard slit spectroscopy routines in
IRAF.  For each night of observations, we coadded the available spectra
to improve the signal-to-noise (S/N) of weak lines, especially He I
$\lambda6678$, before rectifying the mean spectrum to a unit continuum using
line-free regions.  The mean spectrum is shown in Fig.~\ref{fig:meanspec}.

Over the 6 months of our observations, we detected changes in the overall
strength of the H$\alpha$ emission but no significant line profile variations. 
The equivalent width of the H$\alpha$ line, $W_{\rm H\alpha}$, was measured by
integrating over the emission line profile. We use the convention that an
emission line has a negative $W_{\rm H\alpha}$, so the absolute value of
$W_{\rm H\alpha}$ is shown in Fig.~\ref{fig:MW_LC}, panel (e). Although the He
I $\lambda6678$ line has much lower S/N than H$\alpha$, we also measured its
equivalent width, $W_{6678}$. The errors in $W_{\rm H\alpha}$ and $W_{6678}$
are about  5\% and 10\%, respectively, due to noise and continuum placement. 
We compare $W_{6678}$ (multiplied by 100 for better contrast) and $W_{\rm
H\alpha}$ in panel (e) of Fig.~\ref{fig:MW_LC}.  Although there is more scatter
in the $W_{6678}$ measurements, we found that the overall strength of He I
$\lambda6678$ generally tracks the strength of H$\alpha$ well.
Table~\ref{tab:optical} lists the UT date, MJD, time relative to periastron
passage, measured $W_{\rm H\alpha}$, and measured $W_{6678}$ values for each
observation in columns 1--5.


A growth of the equivalent width of the H$\alpha$ line was observed from
$-W_{\rm H\alpha}\simeq62$~\AA\ around 5~days before periastron until $-W_{\rm
H\alpha}\simeq75$~\AA\ about 10~days after it. A decrease in $W_{\rm H\alpha}$ was
observed later, although the poor sampling only allows us to constrain the
start of the decrease between $t_{\rm p} + 18$~days and $t_{\rm p} + 46$~days, when
it was back to $-W_{\rm H\alpha}=62$~\AA\ (note that a baseline level of
$-W_{\rm H\alpha}=54$~\AA\ was measured at apastron by \citealt{distance}). We
interpret the observed behavior as the growth/decay of the mass and size of the
Be star disk, caused by the interactions with the pulsar/pulsar wind. {  Unfortunately  due to data sparsity we don't know the exact date when the  behaviour of the line EW starts to decrease, but if future observations during following periastron passages confirm the
possible coincidence of the decrease of $W_{\rm H\alpha}$ with the onset of the
$\gamma$-ray flare, then it could point to a possible triggering mechanism of the flare:
an abrupt change in the state of the circumstellar disk of the massive star.} We
explore this possibility in more detail in the Sect.~\ref{disk_evolution}. 

\begin{figure}
\center
\resizebox{1.0\hsize}{!}{\includegraphics[angle=0]{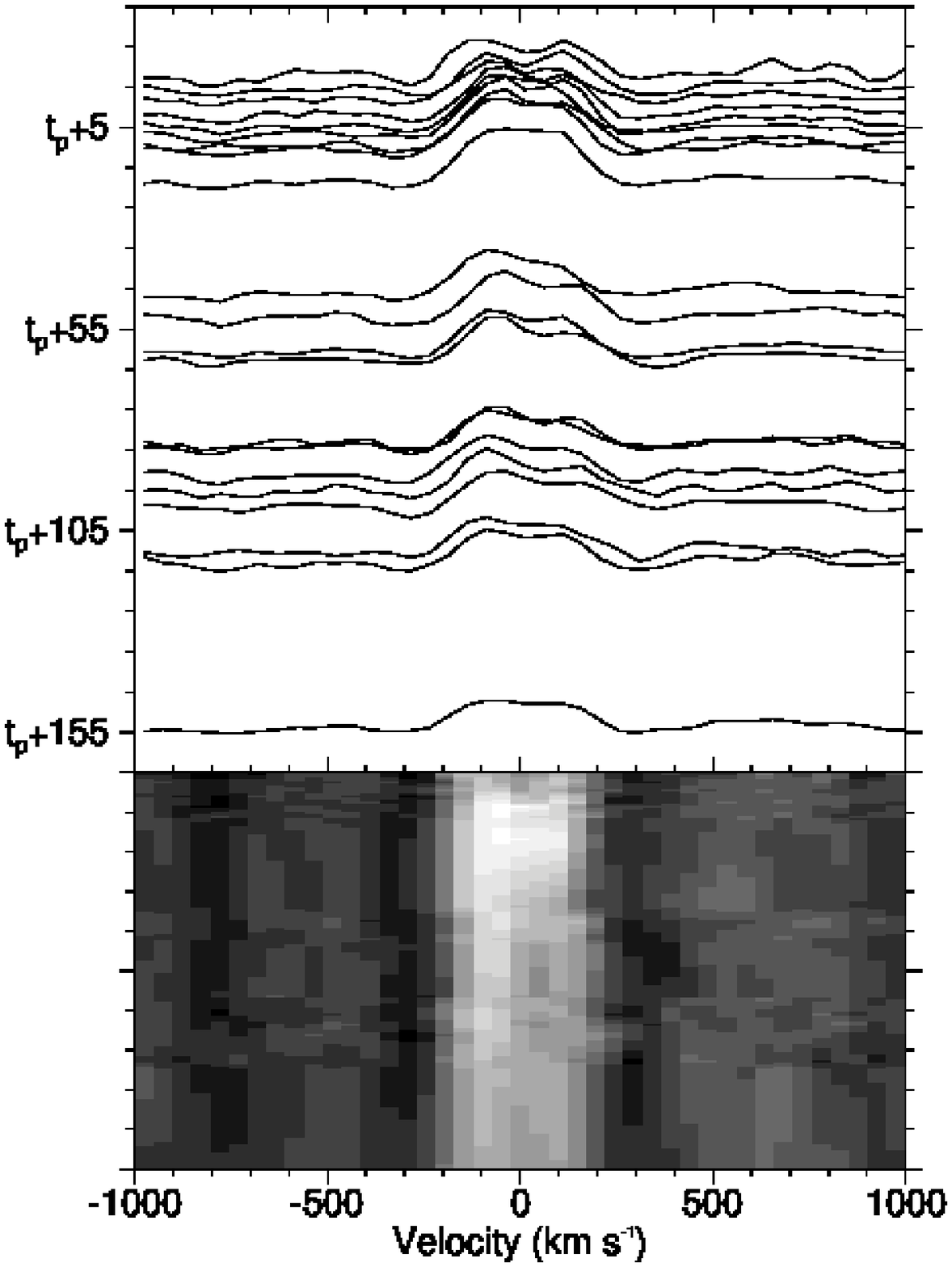}}
\caption{The upper plot shows the He I $\lambda6678$ line profile of LS~2883 over the 6 months of observation, sorted by MJD, and the lower plot shows a gray-scale image of the same line.  The intensity at each velocity in the gray-scale image is assigned one of 16 gray levels based on its value between the minimum (dark) and maximum (bright) observed values.  The intensity between observed spectra is calculated by a linear interpolation between the closest observed phases.}
\label{fig:he6678gray} 
\end{figure}

Figure~\ref{fig:he6678gray} shows the evolution of the He I $\lambda6678$ line
profile.  Despite the low resolving power of our observations, in most of our
spectra the line is resolved with a clear double-peaked shape. The profile is
symmetric within the limits of the S/N during the initial period up to $\sim
t_{\rm p} + 20$~days.  At later times, the line shows a clear excess in the blue
side (negative velocities) of the line relative to the red. 


\begin{table*}
\begin{center}
\caption{Journal of 2010-2011 optical spectroscopic observations of \psrb\ and properties of the Be circumstellar disk of LS~2883 (see Sect.~\ref{disk_evolution} for details).}
\label{tab:optical}
\begin{tabular}{clrcccccc}
\hline
Date &MJD   & $t-t_{\rm p}$   & $W_{\rm H\alpha}$  &  $W_{6678}$  &  $r$ &  $\rho_0$  & $R_{\rm disk}/R_\star$   & $M_{\rm disk}$   \\
&(days) & (days)	&	(\AA) 	     &   (\AA)            &  ($R_\star$)   &  ($10^{-11}$ g~cm$^{-3}$)  &                         &  ($10^{-8} M_\odot$)  \\
\hline
2010-12-05&55535.362  &$-9$&  $-62.3$  &$ -0.44  $&$  33.6$ & 2.82  &   8.8  &  1.31  \\
2010-12-09&55539.341  &$-5$&  $-62.0$  & $-0.39$  &  $28.1$ & 2.79  &  9.1 &   1.16  \\
2010-12-12&55541.997  &$-3$&  $-64.5$  & $-0.39$  &  $25.7$ & 3.03  &   9.5 &   1.19  \\
2010-12-15&55545.304  &$1$&  $-66.4$  & $-0.46$  &  $24.9 $& 3.20  &   9.7 &   1.24  \\
2010-12-18&55548.312  &$4$&  $-70.0$  & $-0.52$  &  $26.4 $& 3.56  &  10.0 &   1.43 \\
2010-12-20&55549.997  &$5$&  $-73.0$  & $-0.48$  &  $28.1 $& 3.86  &  10.2 &   1.61  \\
2010-12-23&55553.305  &$9$&  $-74.8$  & $-0.55$  &  $32.5$ & 4.05  &  10.1 &   1.85  \\
2010-12-24&55554.342  &$10$&  $-74.4$  &$ -0.52$  & $ 34.0$ & 4.00  &  10.0 &   1.88  \\
2011-01-01&55562.319  &$18$&  $-74.7$  &$ -0.56$  &  $47.4 $& 4.04  &   9.3 &   2.31  \\
2011-01-29&55590.226  &$46$&  $-61.6$  &$ -0.40$  &  $91.2 $& 0.60  &   7.7 &   0.50  \\
2011-02-04&55596.000  &$52$&  $-59.7$  &$ -0.45$  &  $99.2$ & 0.58  &   7.5 &   0.51  \\
2011-02-12&55604.290  &$60$&  $-58.0$  &$ -0.43$  &  $110.0$ & 0.44  &   7.4 &   0.39  \\
2011-02-14&55606.307  &$62$&  $-57.2$  &$ -0.42$  &  $112.6$ & 0.44  &   7.3 &   0.38  \\
2011-03-07&55627.231  &$83$&  $-56.6$  &$ -0.35$  &  $137.3$ & 0.43  &   7.3 &   0.38  \\
2011-03-08&55628.002  &$83$&  $-56.3$  &$ -0.37$  &  $138.2$ & 0.43  &   7.3 &   0.38  \\
2011-03-15&55635.002  &$90$&  $-57.2$  &$ -0.43$  &  $145.7$ & 0.44  &   7.3 &   0.38  \\
2011-03-19&55639.002  &$94$&  $-57.2$  &$ -0.42$  &  $149.9$ & 0.44  &   7.3 &   0.38  \\
2011-03-23&55643.163  &$98$&  $-56.2$  &$ -0.42$  & $ 154.2$ & 0.43  &   7.2 &   0.38  \\
2011-04-03&55654.003  &$109$& $ -55.4$ &$ -0.42$  &  $164.9$ & 0.42  &   7.2 &   0.37  \\
2011-04-06&55657.003  &$112$& $ -54.0$  &$ -0.38$  &  $167.8$ & 0.42  &   7.1 &   0.37  \\
2011-05-17&55698.004  &$153$& $ -54.6$  &$ -0.31$  &  $203.3$ & 0.42  &   7.1 &   0.37  \\
\hline
\end{tabular}
\end{center}
\end{table*}

\section{X-ray Observations and Results} \label{x-rays}

We conducted an X-ray monitoring campaign on \psrb\ with \xmm\, \swift\ and
\szk\ telescopes, covering the period between $t_{\rm p}-131$~days and $t_{\rm
p}+79$~days. These observations are summarized in Tables~\ref{xmmdata}, \ref{Swdata}
and \ref{Szdata}. Each of these tables lists an identifier for the data set, UT
date, MJD, time relative to periastron passage, true anomaly of the orbit, and
exposure time for each observation.


\begin{table*}
\begin{center}
\caption{Journal of 2011 \xmm\ observations of \psrb. \label{xmmdata}}
\begin{tabular}{cccccccc}
\hline
Data& Date & MJD& $t-t_{\rm p}$&$\phi$&Exposure&f$_{\rm mos1}$&f$_{\rm mos2}$ \\
 Set&      & (days)&  (days)&(deg)&(ks)&&         \\
\hline
X14&  2011-01-06 & 55567.7 &22 & 86.14&  13.49& $ 1.020\pm{0.007}$&$0.967\pm{0.007}$ \\
X15&  2011-02-02 & 55594.8 &49 &112.26& 13.58& $ 1.05\pm{0.01}$&$ 1.05\pm{0.01}$ \\
X16&  2011-03-04 & 55624.2 &79 &278.57&  8.10& $ 1.02\pm{0.02}$&$ 1.05\pm{0.02}$ \\
\hline
\end{tabular}
\end{center}
\end{table*}

\subsection{\xmm\ observations}

The log of the \xmm\ data analyzed in this paper is presented in
Table~\ref{xmmdata}. In all \xmm\ observations, the source was observed with
the European Photon Imaging Cameras (EPIC) MOS1, MOS2 \citep{herder01} and PN \citep {struder01} detectors in the small window mode with a medium filter.
 The \xmmsp
Observation Data Files (ODFs) were obtained from the online Science
Archive\footnote{http://xmm.vilspa.esa.es/external/xmm\_data\_acc/xsa/index.shtml}
and  analyzed with the Science Analysis Software ({\sc sas}) v11.0.0. During
the data cleaning, all events that have energies above 10~keV and a count rate
higher than 0.5 cts~s$^{-1}$ for the PN or 0.35 cts~s$^{-1}$ for the MOS
detectors were removed.

The event lists for spectral analysis were extracted from a
22.5$^\prime$$^\prime$ radius circle for the MOS1 and MOS2 observations and
from a 45$^\prime$$^\prime$ radius circle for the PN observations. For the
spectral analysis, we made a simultaneous fit to the MOS1, MOS2 and PN data
without imposing constraints on the intercalibration factors. The values
of the MOS1 and MOS2 intercalibration factors f$_{\rm mos1}$, f$_{\rm mos2}$
relative to the PN are given in the last two columns of Table~\ref{xmmdata}.


\subsection{\swift\ observations}

\swift\ has closely monitored the \psrb\ 2010 periastron passage, and the data
log is shown in Table~\ref{Swdata}. The Swift/X-ray Telescope (XRT) \citep{gehrels04} data were taken in photon mode with
$500\times500$ window size. We processed all the data with standard procedures 
using the FTOOLS\footnote{http://heasarc.gsfc.nasa.gov/docs/software}  task
\texttt{xrtpipeline} (version 0.12.6 under the HEAsoft package 6.12).  We
extracted source events from a circular region with a radius of 1$^\prime$, and
to account for the background  we extracted events from a nearby circle of the
same radius.  Due to the low countrate (less than 0.4 cts~s$^{-1}$),  no pile
up correction was necessary.

\begin{table}
\begin{center}
\caption{Journal of 2010-2011 \swift\ observations of \psrb. \label{Swdata}}
\begin{tabular}{c c c r c c}
\hline
Data& Date & MJD& $t-t_{\rm p}$&$\phi$&Exposure \\
 Set&      & (days)   &  (days)&(deg)&(ks)         \\
\hline
Sw5&  2010-08-06 & 55414.3 &$-131$ &210.6&  2.98 \\
   &  2010-08-08 & 55416.0 &$-129$ &210.8&  3.29 \\
   &  2010-08-09 & 55417.4 &$-128$ &211.1&  3.34 \\
   &  2010-08-12 & 55420.0 &$-125$ &211.5&  2.41 \\
   &  2010-08-15 & 55423.2 &$-122$ &212.0&  3.19 \\
Sw6&  2010-11-20 & 55520.2 &$-25$ &255.2&  3.87 \\
Sw7&  2010-11-25 & 55525.6 &$-19$ &264.5&  4.18 \\
Sw8&  2010-11-30 & 55530.5 &$-14$ &276.5&  3.36 \\
Sw9&  2010-12-14 & 55544.6 &$-1$ &355.6&  4.30 \\
Sw10&  2011-01-19 &55580.7 &$35$ &117.0&  3.98 \\
Sw11&  2011-01-20 &55581.0 &$36$ &117.2&  4.17 \\
Sw12&  2011-01-22 &55583.7 &$38$ &119.5&  4.13 \\
\hline
\end{tabular}
\end{center}
\end{table}

After spectral extraction, the data were rebinned with a minimum of 25  counts
per energy bin to allow $\chi^2$ fitting. The anciliary response file was
generated with \texttt{xrtmkarf}, taking into account vignetting and 
point-spread function corrections. In our analysis we used the
\texttt{swxpc0to12s6\_20070901v011.rmf} spectral  redistribution matrix for
observations Sw5, Sw10, Sw11 and Sw12. For Sw6, Sw7, Sw8, Sw9
\texttt{swxpc0to12s6\_20010101v013.rmf} was used instead.

The first set of \swift\ observations was taken in August 2010, when the
source flux was rather weak. After checking that the spectrum of the source
varies only within the error limits during these observations, we combined all
the \swift\ August data in order to have better statistics (Sw5 observation).


\subsection{\szk\  observations}

There were three  \szk\  observations taken shortly after the  2010  periastron
passage of \psrb\ (see  Table~\ref{Szdata}).  The  \szk\  observations were
performed with three X-ray CCDs  (the X-ray Imaging Spectrometers or XISs;
\citealt{koyama07}) in the range 0.3--12~keV and the Hard X-ray Detector
\citep[HXD; ][]{takahashi07} in the range 13--600~keV. The XISs consists of one
back-illuminated CCD camera (XIS-1) and two front-illuminated CCDs  (XIS-0 and
XIS-3). In this paper we discuss only the results of the XIS observations,
while the HXD data will be discussed in a separate paper. Our analysis was
performed using the HEAsoft software package (version 6.12).

\begin{table}
\begin{center}
\caption{Journal of 2011 \szk\  observations of \psrb. \label{Szdata}}
\begin{tabular}{c c c c c c}
\hline
Data& Date & MJD& $t-t_{\rm p}$&$\phi$&Exposure \\
 Set&      & (days)   &  (days)&(deg)&(ks)       \\
\hline
Sz9&2011-01-05&55566.8&22&99.6 &90.0\\
Sz10&2011-01-24&55585.6&41&121.0 &40.3\\
Sz11&2011-02-02&55594.2&49&126.5 &21.5\\
\hline
\end{tabular}
\end{center}
\end{table}

We analyzed the XIS data using pipeline processing version 2.5.16.29.  We made
use of cleaned event files of the XIS observations, with standard event
screening applied. The standard screening procedures include event grade
selections and the removal of time periods such as the spacecraft passage
through the South Atlantic Anomaly, intervals of small geomagnetic cutoff
rigidity, and times of low elevation angle.  We found that XIS-0 gives somewhat
lower $N_{\rm H}$ than others,  which is most likely caused by imperfect
calibrations due to the recently-changed behavior of contamination buildup. 
Thus we report the results from XIS-1 and XIS-3 only.  A joint spectral fitting
of XIS-1 and XIS-3 was done in the energy range of 0.4--9.5~keV. 


\subsection{The X-ray lightcurve}

Panel (c) of Fig.~\ref{fig:MW_LC} shows the X-ray lightcurve of the system. 
Observations made with different instruments at close orbital phases are
consistent with each other, demonstrating good intercalibration. 

From data obtained at orbital phases similar to the archival observations of
previous periastron passages, one can see that the system orbital lightcurve is
stable over several year timescales.  The stability of the orbital lightcurve
allows us to use old and new data simultaneously while analyzing the orbital
evolution of the flux. The sharp rise of the X-ray flux at $t \sim t_{\rm
p}-25$, probably related to the pulsar entering the Be star disk, in 2010 was
closely followed by Swift (observations Sw6 - Sw8).  During this time, the flux
increased by  a factor of 4 in nine days (note that due to the sparse
observations, we probably missed the maximum flux). As the pulsar moved towards
the periastron passage, the flux gradually decreased,  reaching a local minimum
that is only half the peak value near the periastron. Afterwards, we followed
the system through its second passage near the disk, when we again observed an
increase in flux to a value approximately 1.4 times higher than the maximum
value observed during the first disk crossing. After that, the X-ray flux
gradually decreased in a good agreement with previous observations, showing no
unusual behavior during the time of the GeV flare.

\subsection{Spectral Analysis}

\begin{table*} 
\begin{center}
\caption{Spectral parameters for 2010-2011 observations of \psrb.\label{summary}}  
\begin{tabular}{crccccc}
\hline
Data& $t-t_{\rm p}$&$F$(1--10~keV) &$\Gamma$ &$N_{\rm H}$&$\chi^2$(ndof)\\ 
Set&  (days)            &($10^{-11}$ erg cm$^{-2}$ s$^{-1}$)         &&(10$^{22}$ cm$^{-2}$)& \\ 
\hline
Sw5 & $-127$ &$0.17^{+0.01}_{-0.02}$&$1.8\pm{0.09}$&$ 0.5$&8.3 (14)\\
Sw6 & $-25$ &$0.69^{+0.08}_{-0.16}$&$0.99^{+0.13}_{-0.23}$&$ 0.5$&3.6 (7)\\
Sw7 &$-19$ &$1.53\pm{0.12}$&$1.33\pm{0.08}$&$ 0.5$&25.48 (24)\\
Sw8 &$-14$ &$2.60^{+0.12}_{-0.16}$&$1.37\pm{0.06}$&$ 0.5$&61.73 (35)\\
Sw9 &$-1$ &$1.72^{+0.11}_{-0.08}$&$1.56\pm{0.05}$&$ 0.5$&47.9 (39)\\
Sz9 &$22$ &$3.56\pm{0.01}$&$1.76\pm{0.01}$&$ 0.56\pm{0.01}$&2177 (2454)\\
X14 &$22$ &$3.25^{+0.01}_{-0.02}$&$1.71\pm{0.01}$&$ 0.48\pm{0.04}$&2220 (2029)\\
Sw10&$35$ &$2.31^{+0.08}_{-0.11}$&$1.51\pm{0.06}$&$ 0.5$&36.1 (41)\\
Sw11&$36$ &$2.19^{+0.09}_{-0.10}$&$1.48\pm{0.06}$&$ 0.5$&44.7 (42)\\
Sw12&$38$ &$2.02^{+0.12}_{-0.05}$&$1.44\pm{0.06}$&$ 0.5$&33.2 (41)\\
Sz10&$41$ &$2.08\pm{0.02}$&$1.53\pm{0.01}$&$ 0.51\pm{0.01}$&1566 (1528)\\
X15 &$49$ &$1.61^{+0.02}_{-0.01}$&$1.45\pm{0.01}$&$ 0.44\pm{0.01}$&1418 (1453)\\
Sz11&$49$ &$1.77^{+0.02}_{-0.04}$&$1.46\pm{0.02}$&$ 0.48\pm{0.01}$&749 (769)\\
X16 &$79$ &$1.13^{+0.02}_{-0.01}$&$1.37\pm{0.02}$&$ 0.42\pm{0.01}$&855(851)\\
\hline
\end{tabular}
\end{center}
\end{table*}

The X-ray spectral analyses were done with NASA/GSFC XSPEC v12.7.1 software
package.  A simple power law with a photoelectric absorption describes the data
well, with no evidence for any line features. In Table~\ref{summary} we present
the results of the three parameter fits to the \xmm, \swift\ and \szk\ data in
the 0.5--10~keV energy range. The uncertainties are given at the $1\sigma$
statistical level and do not include systematic uncertainties. The quality of
the \swift\ data does not allow us to look for both spectral slope and absorption
column density, so we decided to fix the latter to the value $N_{\rm
H}=5\times10^{21}$~cm$^{-2}$ consistent with the value found in the \xmm\ and
\szk\  observations.

Similar to the observations of previous periastron passages, we found that
several months before the periastron the spectral index was quite soft, $\Gamma
\sim 1.8$. It became much harder a month before periastron, as the flux started
to increase. The spectrum softened a bit as the flux reached its maximum, but
it remained harder than $\Gamma = 1.5$. The spectrum softened again as the
pulsar approached periastron and entered the disk for the second time. At the moment when the X-ray luminosity reached its second peak the slope was softer than $\Gamma = 1.7$. As the X-ray luminosity started to decrease the slope remained close to 1.5, except during the very last \xmm\
observation, when the slope became hard again, reaching the value $\Gamma = 0.42$ previously observed at $t-t_{\rm p}=+266$~days \citep{Chernyakova2006}.



\section{\F-\L\ Observations and Results} \label{fermi}

The \F\ \textit{Gamma-ray Space Telescope} was launched on 2008 June 11, from Cape
Canaveral, Florida. The Large Area Telescope (LAT) is an electron-positron pair
production telescope, featuring solid state silicon trackers and cesium iodide
calorimeters, sensitive to photons from $\sim 20$~MeV to greater than $300$~GeV
\citep{Atwood2009}. Relative to earlier \g-ray missions, the LAT has a
large  $\sim 2.4$ sr field of view, a large effective area ($\sim 8000$~cm$^2$
for $>1$~GeV on axis) and improved angular resolution
(PSF is better than $1^\circ$ for 68\% containment at 1~GeV). The \F\ survey
mode operations began on 2008 August 4. In this mode, the observatory is rocked
north and south on alternate orbits to provide more uniform coverage so that
every part of the sky is observed for about 30 minutes every 3 hours.

The analysis of the \F-\L\ data in this paper is similar to the method used in
\cite{Abdo2011_b1259}. We used the P7SOURCE data set and the associated P7SOURCE\_V6 instrument response functions. 
The analysis was performed using the \F\ Science Tools 09-27-01
package \footnote{FSSC is available online at http://fermi.gsfc.nasa.gov/ssc/data/analysis/software/}. A zenith angle cut of $< 100^\circ$ was applied to the data to reject
atmospheric \g-rays from the Earth's limb. The standard binned maximum
likelihood analysis was performed on events extracted from a $20^\circ \times
20^\circ$ region around the location of \psrb\ in the energy range
0.1--300\,GeV. The source model used in the maximum likelihood analysis
included 2 years (2FGL) \F\ catalog \citep{Nolan12} \g-ray sources within the analysed region,
\psrb\, and the diffuse Galactic and extragalactic components.  Galactic
diffuse emission was modelled using the \texttt{gal\_2yearp7v6\_v0.fits} model,
and the isotropic diffuse component was modelled using the
\texttt{iso\_p7v6source.txt} model. The \L\ data analysed in this paper covers
the 4.5 years period between 2008 August 4 and 2013 February 19. This covers
the time period over which  the pulsar was near apastron until well after the
passage of the pulsar through the dense equatorial disk of the Be star. 

First we analysed the whole 4.5~years of available data, modelling the spectrum
of each point source with a catalog model. The normalizations and indexes of
catalog point sources as well as the normalization of the extragalactic diffuse
background were then fixed to their best-fit values in order to build the
\psrb\ lightcurve around the periastron. The power law index of \psrb\ was
fixed to the best-fit value ($2.86$) obtained for the time period around the
periastron passage. Afterwards, the normalizations of \psrb, Galactic diffuse
background and the sources marked as variable in 2FGL catalog (within $5^\circ$
from \psrb) were left free, similar to the procedure used by
\citet{Abdo2011_b1259}.

Panel (b) in Fig.~\ref{fig:MW_LC}  shows the \L\ flux light curve. All data points have a significance higher than 2 $\sigma$. The upper limits on the figure corresponds to 95\% confidence limit. The
time binning was selected to be 7~days for the period of the flare and 14~days
otherwise. The light curve shows upper limits until $t_{\rm p}-25$~days, detections
around a flux level of $0.5\times10^{-6}$~cm$^{-2}$~s$^{-1}$ between $t_{\rm p}-25$~days and
$t_{\rm p}+16$~days, and the GeV flare that reaches a flux level of
$2\times10^{-6}$~cm$^{-2}$~s$^{-1}$ at $t_{\rm p}+30$~days and decays in a roughly linear way,
with the last detection at a flux level of $0.5\times10^{-6}$~cm$^{-2}$~s$^{-1}$ between
$t_{\rm p}+65$~days and $t_{\rm p}+80$~days. The results obtained in this paper are consistent with those in the earlier work by \citet{Abdo2011_b1259}.



\section{\H\ Observations and Results} \label{hess}

The High Energy Stereoscopic System (H.E.S.S.) is an array of  imaging
atmospheric Cherenkov telescopes located in the Khomas Highland of Namibia at
an altitude of 1800 m above sea level.  H.E.S.S.~phase~I (consisting of four
13\,m diameter telescopes) is  optimized for the detection of very  high energy
$\gamma$-rays in the range of 100~GeV to 20  TeV. The total field of view is
$5^{\circ}$ and the angular  resolution of the system is $\lesssim
0.1^{\circ}$. The  average energy resolution is about $15\%$. The H.E.S.S.~I 
array is capable of detecting point sources with a flux  of $1\%$ of the Crab
nebula flux at a significance  level of $5\sigma$ in 25 hours when observing at
low zenith  angles \citep{HESSCrab2006}.

Observations of \psrb\ around its 2010 periastron  passage resulted in a rather
small dataset compared  to observations around the  2004
\citep{Aharonian_B1259_2004_pass} and 2007 \citep{psrb1259_hess09}  periastron
passages. The collected data correspond to  a livetime of about 6\,h
\citep{psrb1259_hess13}. Observations  were performed over five nights, 2011
January 9/10, 10/11, 13/14, 14/15  and 15/16, which corresponds to the period
from $t_{\mathrm{p}}+26$  to $t_{\mathrm{p}}+32$ with respect to the time of
periastron. 

The source was detected at the 11.5 $\sigma$ level \citep{lima}  at a position
compatible with previous \H\  observations. The differential spectrum of the
source follows  a simple power-law with a flux normalization at 1~TeV $N_{0} = 
(1.95\pm0.32_{\mathrm{stat}}\pm 0.39_{\mathrm{syst}})\times10^{-12}$ 
TeV$^{-1}$~cm$^{-2}$~s$^{-1}$ and spectral index $\Gamma = 2.92\pm
0.30_{\mathrm{stat}} \pm 0.20_{\mathrm{syst}}$. The integral flux  above 1~TeV
averaged over the entire observation period is 
$F(E>1\mathrm{~TeV})=(1.01\pm0.18_{\mathrm{stat}} \pm 0.20_{\mathrm{syst}})
\times10^{-12}$ cm$^{-2}$~s$^{-1}$ \citep{psrb1259_hess13}. The  nightly
lightcurve over the observation period (panel (a) in Fig.~\ref{fig:MW_LC}) is
compatible with a constant flux and does not show any hint  of source
variability \citep{psrb1259_hess13}.


Both the flux level and spectral shape from 2011 are in a very good agreement 
with results obtained for previous periastron passages. Moreover, the
comparison of results obtained in 2011 and 2004 observations, which were taken
at similar orbital phases, provides a stronger evidence of the repetitive
behavior of \psrb\ (see Fig.~\ref{fig:MW_LC}) \citep{psrb1259_hess13}. Such
comparison was not possible using the 2004 and 2007 data sets as observations
were performed at different phases before and after periastron.

\H\ observations provide a three-day overlap ($t_{\mathrm{p}}+30$; 
$t_{\mathrm{p}}+32$) in time with the GeV flare detected by \F.  This allows
the direct study of a possible flux enhancement  in the TeV band over the
timescale of the GeV flare. A careful  statistical study showed no evidence of
any significant flux  enhancement, which leads to the conclusion that the GeV
flare emission  is of a different nature than the TeV emission
\citep{psrb1259_hess13}.

\section{Discussion} \label{discussion}

The broad-band non-thermal emission from the PSR~B1259$-$63 system is produced
in the interaction of the relativistic pulsar wind with the stellar wind of the
massive Be star. The high eccentricity of the orbit is responsible for the
episodic nature of the source activity, with bright outbursts occurring during
the periods of the pulsar's close passage near the massive star. The
characteristic "double peak" structure of the orbital lightcurves of the source
in the radio, X-ray and, possibly, TeV $\gamma$-ray bands is naturally
interpreted as occurring due to the inhomogeneity of the stellar outflow of the
fast-rotating Be star. The peaks of the non-thermal emission are associated
with the periods of the pulsar's passage through the denser and slower gas in
the equatorial disk of the Be star. This behavior has also been observed during
the previous periastron passages and is clearly seen in the radio and X-ray
bands also during the 2010-2011 periastron passage, shown in
Fig.~\ref{fig:MW_LC}.  

The orbit-to-orbit lightcurves in the radio band, which are qualitatively
similar with the maxima occurring at similar orbital phases, exhibit up to
$\sim 50\%$ differences in the overall flux level between different periastron
passages \citep{Johnston2005}. Contrary to the radio lightcurve, the behavior
of the source in the X-ray band is remarkably stable. Measurements of the X-ray
flux at the same orbital phases, spaced by several 3.4~yr orbital periods,
reveal nearly the same flux. 

\F\ LAT observations of the 0.1--10~GeV $\gamma$-ray activity of the source
during the 2010-2011 periastron passage have revealed a puzzling flaring
activity, which occurred after the periastron passage and also after the
post-periastron transit of the equatorial disk of the Be star by the pulsar.
The GeV flaring activity was remarkable in the sense that the energy output of
the source in this band reached a theoretical maximum given by the spin-down
luminosity of the pulsar, $8\times 10^{35}$ erg~s$^{-1}$, assuming isotropic
emission in this energy band.  \citet{Abdo2011_b1259} have noticed that the GeV
band flare was peculiar in the sense that it had no obvious counterparts at
lower energies, in the radio and X-ray bands. 

The broad-band observations of the source reported here generally support the
conclusion of \citet{Abdo2011_b1259} on the ``orphan'' nature of the GeV flare.
The only data that could reveal an irregularity in the source behavior
coincident in time with the moment of the GeV flare are the optical
spectroscopic data shown in panel (e) of Fig.~\ref{fig:MW_LC}. These data show
a clear decrease of the equivalent width of the H$\alpha$ line that could
coincide with the GeV flare. A gap in the optical spectroscopy data during the
period 18--46~days after the periastron does not allow us to tell if the
decrease of $W_{\rm H\alpha}$ is exactly coincident with the moment of the GeV
flare (30 days after the periastron). However, it seems that the decrease of
the line strength is delayed with respect to the start of the radio outburst
that happened 15 days after the periastron. The peculiar behavior of the
H$\alpha$ line and its possible relation to the GeV flare deserve special
attention.

\subsection{Evolution of the Be star disk mass and size based on the H$\alpha$ line measurements \label{disk_evolution}}


With multiple observations of $W_{\rm H\alpha}$ available, we can estimate the
changing size and mass of the circumstellar disk through the periastron
passage.  \citet{grundstrom2006} describe a simple model to measure the ratio
of the projected effective disk radius (the radius at which half of the
H$\alpha$ emission originates) to the stellar radius, $R_{\rm
disk}$/$R_{\star}$, and the density at the base of the disk, $\rho_0$, using
$W_{\rm H\alpha}$, the stellar effective temperature $T_{\rm eff}$, and disk
inclination angle $i_{\rm disk}$ as input parameters.


\citet{distance} found that LS~2883 is highly distorted due to its rapid
rotation, with a polar radius (or equivalent non-rotating radius) $R_\star =
8.1 \; R_\odot$ and a bulging equatorial radius $R_{\star, \rm eq} = 9.7 \;
R_\odot$.  The equator is significantly cooler than the poles (27500 K vs.\
34000 K) due to gravity darkening. They also found that the Be star is inclined
with $33^\circ$ and has an angular velocity $\Omega$ of about 88\% of the
critical rotation value at which the centrifugal force at the equator equals
the gravitational force. Using a Roche model for such a rapidly rotating star
\citep{maeder2009}, we find that the mean temperature averaged over the stellar
surface is about 30200 K, so we define $T_{\rm eff}$ accordingly.


For LS~2883, the Be disk should be highly truncated near periastron, both due
to the gravitational influence of the pulsar (which is observed in other Be
binaries using long baseline optical interferometry; \citealt{gies2007}) and
due to disruption by the pulsar wind ram pressure (predicted by simulations of
\citealt{Okazaki2011, Takata2012}). The truncation distance expands rapidly
after periastron passage. Therefore we used the orbital solution of
\citet{Wang2004}, a stellar mass of 31 $M_\odot$ \citep{distance}, and the
typical neutron star mass of 1.4 $M_\odot$ to calculate the binary separation
distance, $r$, as a function of time.  The separation values range from $24.8
\; R_{\star} < r < 203.3 \; R_\star$ over the course of our observations. We
fixed the outer disk boundary to $r$ when the stars are close ($r < 100 \;
R_\star$), and we used an outer boundary of 100 $R_\star$ in accordance with
the recommendation of \citet{grundstrom2006} for times when $r > 100 \; R_\star$.


To estimate the total mass of the disk, we used an axisymmetric, isothermal
density distribution,
\begin{equation} \label{densityeqn}
\rho(r,z) = \rho_0 \left (\frac{R_\star}{r} \right )^n \exp \left[-\frac{1}{2} 
\left(\frac{z}{H(r)}\right)^2 \right]
\end{equation}
\citep{carciofi2006} and a radial density exponent $n = 3$, typical of other Be star disks \citep{gies2007}.  The scale height of the disk is 
\begin{equation}
H(r) = H_0 \left( \frac{r}{R_{\star}}\right)^\beta,
\end{equation}
where
\begin{equation}
H_0 = \frac{a}{V_{\rm crit}} R_{\star},
\end{equation}
\begin{equation}
a = \sqrt{\frac{kT}{\mu m_{\rm H}}},
\end{equation}
and $\beta = 1.5$ for an isothermal disk \citep{bjorkman2005, carciofi2006}. 
We integrated this density distribution from the equatorial stellar surface at
$R_{\star, \rm eq}$ out to the disk truncation radius, described above, to
estimate the total disk mass, $M_{\rm disk}$.

The resulting $\rho_0$, $R_{\rm disk}/R_\star$, and $M_{\rm disk}$ are listed
in Table~\ref{tab:optical}. We emphasize that these disk measurements should be
viewed with caution since our assumption of an axisymmetric, isothermal disk is
overly simplistic. Deviations from this simple disk structure may produce order
of magnitude variations in the calculated mass, and these effects are discussed
more throughly by \citet{mcswain2008}. { A further source in error for our disk mass measurement is the truncation radius assumed in our model.  For a pulsar orbit that is inclined 90 degrees relative to the Be disk, the separation distance when the pulsar becomes eclipsed by the disk, 42.2 $R_\star$, is slightly higher than the periastron distance.  This implies slightly higher disk masses near periastron.  However, if the disk is truncated instead at the star's effective Roche lobe radius, then we predict slightly lower disk masses.  These revised masses are well within the order of magnitude error that is inherent to the assumptions of our model.}

Within our model we found that the Be disk grew in mass as the binary went
through periastron passage in 2010 December.  Tidal disruption by the neutron
star may have triggered this sudden growth in disk { mass}. \citet{moreno2011}
find that the rate of energy dissipation over the stellar surface reaches a
maximum amplitude near or slightly after periastron passage in very eccentric
binaries, prompting an increase in stellar activity near that orbital phase. 
Other Be stars (eg.\ $\delta$ Sco; \citealt{Miroshnichenko2001,
Miroshnichenko2003}) have been observed to exhibit disk outbursts around the
time of periastron passage, so the disk growth of LS~2883 is not unusual.  

The times of GeV flaring in \psrb~correspond to an epoch of disk reduction.  
We speculate that the GeV flaring could be due to the interaction of a mass
stream being pulled away from the disk to collide with the relativistic
pulsar's wind.  

Additional information on the properties of the disrupted disk
could be found from the  He I $\lambda6678$ line profile, shown in
Fig.~\ref{fig:he6678gray}.  Despite the low resolving power of our
observations, in most of our spectra the line is resolved with a clear
double-peaked shape.  During the disk growth period, the peaks are symmetric
within the limits of the S/N.  However, during the disk reduction period, the
He I $\lambda6678$ line shows a clear excess in the blue side of the line
(negative velocities) relative to the red one.  The asymmetry is sustained for
at least $\sim60$ days, far longer than the expected 1--2 days required for
disk material to circle the star.  The asymmetry may indicate a slowly moving
spiral density wave in the disk, typical of density waves observed in other Be
stars \citep{porter2003}.  The pulsar wind probably does not significantly
alter the ionization levels in the circumstellar disk.  Since He I will ionize
at a lower temperature than H I, it would be expected to be fully ionized near
periastron if the pulsar wind's influence on the disk produces strong
ionization effects.  However, we observe that the H$\alpha$ and He I
$\lambda$6678 line strengths track each other well in Fig.~\ref{fig:MW_LC}, and
no significant deviations are found within the limits of noise. The
similarities with other Be stars in binary systems, which suffer gravitational
disruption, and the non alteration of the disk ionization, suggest that the
disk disruption seen in the optical data might simply be driven by gravitational tidal forces instead of a pulsar interaction.  


\subsection{The radio outflow during the GeV flare}

The high-energy particle outflow from the system is unresolved in most of the
energy bands except for the radio. Both previously reported high-resolution
observations in the radio band \citep{Moldon11} and the observations  reported
here reveal the extended nature of the radio source, about 50~mas in size. At a
distance $d\simeq 2$~kpc \citep{distance}, this corresponds to a projected
linear size of the source $R\simeq 1.5\times 10^{15}\left[d/2\mbox{
kpc}\right]$~cm, which is two orders of magnitude larger than the orbital
separation at periastron. Our analysis of the high-resolution LBA image
obtained 29 days after the 2010 periastron passage has two important
implications.

Firstly, we have found a radio structure that is very similar to the one found
21 days after the 2007 periastron passage (see Fig.~1-middle in
\citealt{Moldon11}). This confirms the presence of an extended structure in
each periastron passage and their similar appearances, albeit with a slightly
more negative position angle in the image presented here. This may be a
consequence of the different orbital phases of the observations, and it will be
discussed in a future paper including four additional LBA observations
conducted around the 2010 periastron passage.

Secondly, the position of the pulsar is determined for the first time with the
same data used to obtain the radio morphology of the nebula extending towards
the North-West, allowing us to make a direct measurement of the pulsar's
position inside the nebula. We found that the pulsar is located around 5~mas
towards South-East of the peak of the radio nebula. This is still marginally
compatible with the predicted pulsar position in the 2007 periastron passage
(see the red cross in Fig.~1-middle in \citealt{Moldon11}), which was an
indirect estimation affected by unmodelled ionospheric uncertainties that also
depended on the uncertain proper motion of the binary system and on the orbital
motion of the pulsar around the massive star. With the relative positions of
the pulsar and the nebula obtained here, we find that the overall morphology of
the source is consistent with a cometary tail extending behind the pulsar.

The projected displacement of the peak of the nebular emission from the pulsar
position, $R_{\rm r}\simeq 1.5\times 10^{14}\left[d/2\mbox{ kpc}\right]\mbox{
cm}$, provides an estimate of the distance scale at which the optical depth
becomes comparable to the free-free absorption, which suppresses the radio
emission from the innermost region.  The measured distance to the radio
emitting region is in a good agreement with the estimate of \citet{zdziarski}
\begin{eqnarray}
&& R_{\rm r}\simeq 2\times 10^{14}\left[\frac{\dot M_{\rm w}}{10^{-7}~M_\odot~\mbox{yr}^{-1}}\right]^{2/3}
\left[\frac{v_\infty}{10^8~\mbox{cm s}^{-1}}\right]^{-2/3}\times \\ \nonumber 
&& \times \left[\frac{\nu}{2.3~\mbox{GHz}}\right]^{-2/3}
 \left[\frac{T}{3\times 10^4~\mbox{K}}\right]^{-1/2} \left[\frac{f}{0.1}\right]^{-1/3}~\mbox{cm}
\end{eqnarray}
for a typical mass loss rate of the Be star $\dot M_{\rm w}$, asymptotic speed,
$v_\infty$, clumping factor, $f$, and temperature, $T$, of the stellar wind.
For \psrb\ the terminal velocity of the wind was estimated to be $1350 \pm 200$
km~s$^{-1}$ \citep{B1259_terminationVelocity}.



The radio emission is produced via the synchrotron mechanism by electrons with
energies \citep{zdziarski}
\begin{equation}
E_{\rm e,r}\simeq 0.1\left[\frac{B_{\rm r}}{10\mbox{ mG}}\right]^{-1/2}\left[\frac{\nu}{1\mbox{ GHz}}\right]^{1/2}\mbox{ GeV}
\end{equation}
where $B_{\rm r}$ is the strength of magnetic field in the radio emission
region. If such electrons are (a) injected during the GeV flare and (b) are
able to escape towards the radio emission region with the speed comparable to the speed of light, they would reach the radio emission region in less than two
hours from the start of the flare. Therefore, the absence of a radio
counterpart of the GeV flare (mostly from archival data at similar orbital
phases) implies that either the 100~MeV electrons are not injected during the
flare or they escape at a speed much lower than the speed of light. 

Slow escape of the radio-emitting plasma has been recently revealed in the
observations of another $\gamma$-ray loud binary system, LS~I~+61~303
\citep{cher12}, in which the speed of the high-energy particle loaded outflow
was found to be comparable to the speed of the stellar wind. Adopting such a
model for the PSR~B1259$-$63 system would imply a time delay $t_{\rm r}=R_{\rm
r}/v_\infty\simeq 17$~days for the radio "echo" of the GeV flare. This time period
was not covered by the ATCA radio observations reported here (see panel (d) of
Fig.~\ref{fig:MW_LC}), so it remains unclear whether the GeV flare was
associated with a delayed radio counterpart or whether electrons with energies
in the $\sim 0.1$~GeV range were injected in the flare. No evidence of such a
flare is present in the archival data (see panel (d) of Fig.~\ref{fig:MW_LC}
for the details of 2004 periastron passage), but at the moment we do not know
whether the GeV flare occurs each periastron passage so we cannot make any firm
conclusions.

\subsection{X-ray and TeV observations}

Similarly to the radio observations, neither X-ray nor the TeV observations
reveal obvious counterparts to the GeV flare. However, contrary to the radio
emission, the X-ray and TeV emission are, most probably, produced directly 
inside the binary orbit so that no time delay of the X-ray and TeV band
emission is expected. 

Although the time coverage of the HESS TeV lightcurve during the 2010-2011
periastron passage is not sufficient to draw definitive conclusions on the
presence/absence of the TeV flaring activity during the whole period of
the GeV band flare, a careful statistical study showed no evidence of any
significant flux enhancement in TeV energy band right at the beginning of
the GeV flare \citep{psrb1259_hess13}. Moreover, one could
notice from the top panel of Fig.~\ref{fig:MW_LC} that the TeV flux
measurements one month after the periastron are consistent with the
previous measurements during the 2004 periastron passage at similar times.
This indicates that the orbit-to-orbit behavior of the source in the TeV
band might be stable, similar to the behavior in the X-ray band.  In fact,
the X-ray and TeV band flux might be produced by one and the same electron
population with $\sim$TeV energies, via the synchrotron and inverse
Compton mechanisms. If this is the case, one could combine the 2004 and
2010-2011 data into an orbit-folded lightcurve. Such a lightcurve would
reveal a broad post-periastron maximum compatible with the post-periastron
maxima of the X-ray and radio lightcurves, but there is no pronounced
maximum occurring at the moment of the GeV band flare  (see Fig.~\ref{fig:MW_LC}).

In the X-ray band, non-simultaneous data from different periastron passages
are compatible with the interpretation that the post-periastron maximum of the lightcurve may have a two-peak
structure (see Fig.~\ref{fig:MW_LC}). Based on the observed trend in the
post-periastron X-ray lightcurve, we should expect the X-ray flux about 30~d
after the periastron to decline. However, instead we see that the 2007 {\it
Chandra} flux at $t_{p}+29$~days is comparable to the  the 2010 measurements by
\szk\ and {\it XMM-Newton} some 22~days after the periastron. Thus, the flux
observed by {\it Chandra} in 2007 at $t_{p}+29$~days might be an X-ray counterpart to
a GeV flare (if the GeV flare is a periodic phenomenon). However, similarly to the case of the TeV observations, it is not
possible to draw definitive conclusions on the presence/absence of the X-ray
counterpart of the GeV flare because of the lack of the systematic monitoring
of the source close to the flare period. 

Repeated observations of the source during the next periastron passage, with a
denser time coverage both in X-rays and in the TeV band, are essential to
clarify the existence of the X-ray and TeV counterparts of the GeV flare as
well as the question of the recurrent nature of the flare.

\subsection{Possible connection of the Be star disk perturbation and the GeV Flare.}

{ In this paper we discuss the possibility that the detected GeV flare could be
related to the reduction in the size/mass of the equatorial disk of the massive
Be star in the system.} Unfortunately, a gap in the H$\alpha$ data does not
allow us to make a firm identification of the start of the GeV flare with the
start of the decrease of the equivalent width of the H$\alpha$ line.  The
possibility of triggering the flare by the disk disruption event has to be
verified with the future observations with denser coverage of the H$\alpha$
measurements around the onset of the flare.

The decrease of $W_{\rm H\alpha}$ and the enhanced blue wing of the He I
$\lambda$6678 line indicate the presence of strong perturbations in the
interacting pulsar wind - stellar wind system. In the absence of perturbation
of the Be star disk, the relativistic particles can escape from the system
along a bow-shaped contact surface of the pulsar and stellar winds and in a
direction opposite to the contact surface apex. A part of the pulsar wind power
emitted not in the direction of the bow shock apex is able to escape to
large distances, like in a typical large-scale pulsar wind nebula. Only the
power emitted in the direction of the bow shock is converted to radiation.

{ In the model discussed in \cite{Kong12} the observed X-ray and GeV emission is explained as synchrotron emission from the postshock relativistic electrons Doppler-boosted at the particular orbital phase. This model is able to describe the observed spectra pretty well, but fails to explain the substantial shift of the X-ray and GeV light curve peaks. 

In the modelling one should take into account that the strong perturbation of the disk destroys the regular bow-shaped contact
surface between the pulsar and stellar outflow.} It is possible that the fly-by
of the disk material near the pulsar destroys the regular geometry of the
relativistic particle and electromagnetic field flow in the unshocked pulsar
wind. Once the magnetic field in the pulsar wind ceases to be aligned with the
particle flow, high-energy particles in the wind immediately lose their energy
via synchrotron radiation.  In such a scenario the energy of electrons
responsible for the GeV flare should be in the 100~TeV range:
\begin{equation}
E_{\rm e,flare}\simeq 10^2\left[\frac{B_{\rm pw}}{1\mbox{ G}}\right]^{-1/2}\left[\frac{\nu}{1\mbox{ GHz}}\right]^{1/2}\mbox{ TeV}
\end{equation}
assuming that magnetic field in the pulsar wind is at the level of $B_{\rm
pw}\sim 1$~G at  distances comparable to the binary separation distance ($\sim
10^{13}$~cm). The energy of electrons responsible for the highest energy
synchrotron emission is in the 100~TeV range, which is close to the PeV
energies of electrons responsible for the recently discovered GeV flares of the
Crab pulsar / pulsar wind nebula system \citep{Abdo2011_crabflare,
Buehler2012}. This suggests a possibility that, in both sources, the flares
could be produced via the same mechanism. In the case of the Crab flares, the
short timescale $t_{\rm crab}<1$~day of the variability suggests a relatively
compact size of the flaring source of $l_{\rm crab}\le 10^2$~AU. This distance
scale is comparable to the size of the extended source revealed by the VLBI
observations of \psrb. Taking this into account, the appearance of 100~TeV
electrons responsible for the \gr\ synchrotron emission in the \psrb\ system
does not appear unreasonable.

In the synchrotron scenario, the duration of the flare is estimated by the time
of the fly-by of the disk material near the pulsar, so that 10--30~days is a
reasonable estimate, assuming a typical speed of the stellar wind and taking
the binary separation distance as the estimate of the size of the region
occupied by the pulsar wind. 

The absence of the flare counterparts in the radio, X-ray and TeV bands could
also be reasonably explained by the high efficiency of synchrotron energy
losses for the 100~TeV particles. Most of the power of the pulsar wind is
converted into the GeV band emission, with only minor fraction of the power
left for emission at much lower energies. Synchrotron cooling of the 100~TeV
particles would form a characteristic $dN/dE\sim E^{-2}$ low-energy tail in the
electron distribution. The spectrum of synchrotron emission from this
low-energy tail would have a slope $dN_\gamma/dE\sim E^{-1.5}$, so that the
power emitted in the X-ray band is some $\sim 3$ orders of magnitude lower than
the power of the GeV band emission. Taking into account that the luminosity  of
inverse Compton emission in the TeV band is comparable to the luminosity of the
X-ray emission, one arrives at a conclusion that no TeV band counterpart of the
flare is expected to be detectable. 

An alternative possibility is that the GeV flaring emission is produced via the
inverse Compton, rather than synchrotron, mechanism. Such a possibility was
considered by \citet{khan12}. In this case, the energies of electrons in the
unshocked pulsar wind are much lower, in the GeV range. A fraction of the
observed GeV emission from the system is thus produced via inverse Compton
scattering of the UV radiation coming either directly from the Be star or from
its circumstellar disk by the unshocked pulsar wind electrons. The fraction of
the GeV flux produced in this way depends on the geometry of the region
occupied by the unshocked pulsar wind. The luminosity of the inverse Compton
emission from the unshocked pulsar wind could strongly increase if the volume
occupied by the unshocked pulsar wind increases. This scenario requires a very
high efficiency of reprocessing stellar radiation in the Be star disk, with up
to a half of the UV luminosity of the system being due to the emission from the
disk, rather than from the star. 

A strong perturbation of the equatorial disk of the Be star could, in
principle, lead to an enhancement of the luminosity of the inverse Compton
emission from the unshocked pulsar wind. Indeed, a natural consequence of the
destruction of the disk is that the volume occupied by the unshocked pulsar
wind rapidly grows. Electrons/positrons in the unshocked pulsar wind propagate
to larger distances and could give away a larger fraction of their energy to
the inverse Compton radiation while remaining in the unshocked wind.  With a
suitable assumption about the UV luminosity of the circumstellar disk { (which turns out to be very high, exceeding the stellar luminosity, possibly as a result of local heating of the Be star disk by the pulsar crossing, see \cite{khan12})}, one could find that nearly 100\% of the spin-down luminosity of the pulsar could be
converted into $\gamma$-ray power in the unshocked wind if its volume becomes 
sufficiently large during the flare. 

A potential problem of such a scenario would be to explain why the inverse
Compton luminosity of the system does not reach the spin-down luminosity of the
pulsar before the destruction of the disk. Indeed, before the disk destruction,
the volume occupied by the unshocked pulsar wind is small so that the
efficiency of the inverse Compton energy loss in the unshocked pulsar wind zone
is low. As a result, the inverse Compton luminosity of the unshocked wind is
much less than 100\% of the spin-down power. However, electrons and positrons
from the pulsar wind do not stop to lose energy via the inverse Compton
emission when they enter the shocked pulsar wind zone at the same rate as in
the unshocked pulsar wind zone. If the unshocked pulsar wind
electrons/positrons release 100\% of their energy into the inverse Compton
emission while escaping from the system during the flaring period, the shocked
pulsar wind electrons/positrons should also release 100\% of their energy via
the same channel while escaping through the shocked pulsar wind before the
flare. Thus, the inverse Compton luminosity of the system is expected to be at
100\% of the pulsar spin-down power throughout the pulsar's passage through the
disk, not only at the moment of the disk destruction.  A possible way out of
this difficulty might be the existence of an alternative non-radiative channel
of energy dissipation of the pulsar wind electrons/positrons in the shocked
pulsar wind region, a question that requires further investigation. 

{Besides the optical photons coming from the star and the disk, X-ray photons produced by the shocked pulsar wind can act as seed photons for the observed GeV emission. Such a possibility was discussed by \cite{Petri11} and \cite{Dubus13}. In \cite{Petri11}  GeV emission is generated rather close to the pulsar, and the X-ray photons are scattered by the relativisitic pairs in the striped pulsar wind. In this case the GeV flare was interpreted by the authors as a lucky combination of the geometry and  a presence of additional seed photons coming from the shocked region. In the \cite{Dubus13} model the GeV emission is due to the IC scattering of the X-ray photons by the schocked relativistic wind. In this model the GeV flare is expected to peak near the inferior conjunction, but the reason of the delay between the X-ray and GeV peak in this model  is not clear.}

{ \cite{Kirk13} interprete the observed flare as IC scattering of the optical photons on the precursor of superluminal waves roughly 30 days after the periastron passage. This model does not require an additional source of photons, but it is unclear what triggers the formation of the precursor well after the exit from the disk. This model also predicts a preperiastron flare, which has not been observed at any wavelength.}

\section{Conclusions} \label{conclusions}

In this paper, we have reported the results of multi-wavelength observations of
\psrb\ during its 2010 periastron passage. This was the first periastron for which complete monitoring in the high-energy gamma-ray band was available. These data have
revealed a puzzling GeV flare occurring $\simeq 30$~days after the periastron
\citep{Abdo2011_b1259}.  Our multi-wavelength data show that  the source
behavior in the radio and X-ray band is qualitatively similar to the previous
periastron passages and that there are no obvious counterparts of the GeV flare
in other wavebands, from radio to TeV $\gamma$-rays. However, both in the
X-rays and radio data from previous periastron passages there might be small
irregularities in the behaviour of the lightcurves at the moment of the onset
of the GeV flare. The possible relation of these irregularities to the GeV
flare has to be verified with simultaneous data during future multi-wavelength
campaign for the next periastron passage. 

The orbital lightcurves in the radio band do exhibit orbit-to-orbit variations.
On the other hand, the X-ray lightcurve is remarkably stable. This might be
related to the fact that the X-ray emission is produced directly inside the
binary system, while the radio emission is produced by the high-energy outflow
reaching distances 10--100 times larger than the binary system size. The
time coverage of the source in the TeV band is not sufficient to judge whether
the source is variable from orbit to orbit.

Optical spectroscopy data reveal the evolution of the H$\alpha$ line strength,
which indicates changes in the state of the circumstellar disk of the Be star
induced by the close passage of the pulsar. The pulsar first induces growth of
the disk in mass and size. The disk growth stops after the periastron passage
and further interaction of the pulsar / pulsar wind with the disk leads to a
perturbation of the disk structure, which possibly triggers the GeV band flare.
This perturbation manifests itself in the reduction of the equivalent widths of
the H$\alpha$  and He I $\lambda$6678 lines.  
On the other hand, the disk disruption that we observe may be a normal occurrence in binary Be systems due to gravitational interactions near the close passage.  
Unfortunately, the optical
spectroscopy data have a gap at the moment of the onset of the GeV flare, so we
are not able to make an unambiguous link between the flare onset and changes in
the Be star disk state. This possible link should be verified with a denser
optical spectroscopy coverage of the periastron passage in 2014. 

We have measured for the first time the position of the pulsar and the unpulsed
extended emission from the same data set in the radio band. Such a measurement
removes the systematic uncertainty related to the cross-calibration of
different data sets which has affected previous measurements. This measurement
pinpoints the position of the binary system within an extended radio emission
region of the size $\sim 100$~AU. We find that the overall morphology of the
radio source is compatible with a ``cometary tail'' extending behind the
pulsar. 

There are a number of open questions that should be addressed in new
multi-wavelength observations during the next periastron passage in 2014. In
particular, it is not clear whether the GeV flare is recurrent and, if it is,
whether it occurs at a particular orbital phase. The triggering mechanism of
the flare has to be clarified. We should investigate more deeply possible
multi-wavelength signatures of the triggering mechanism of the flare with new
simultaneous observations densely covering the time span of the flare.  Our
results point to a possible relationship between the GeV flare and changes in
the Be star disk state. These changes could hopefully be traced by the
variations of the strength of emission lines from the disk or of its column
density.

\section{Acknowledgements}

The $Fermi$ LAT Collaboration acknowledges support from a number of agencies
and institutes for both development and the operation of the LAT as well as
scientific data analysis. These include NASA and DOE in the United States,
CEA/Irfu and IN2P3/CNRS in France, ASI and INFN in Italy, MEXT, KEK, and JAXA
in Japan, and the K.~A.~Wallenberg Foundation, the Swedish Research Council and
the National Space Board in Sweden. Additional support from INAF in Italy and
CNES in France for science analysis during the operations phase is also
gratefully acknowledged.

The Parkes radio telescope and the Australian Long Baseline Array are part of
the Australia Telescope which is funded by the Commonwealth Government for
operation as a National Facility managed by CSIRO. We thank our colleagues for
their assistance with the radio timing observations. This work made use of the
Swinburne University of Technology software correlator, developed as part of
the Australian Major National Research Facilities Programme and operated under
licence.  

A.N. is grateful to the Swiss National Science Foundation for the support under grant PP00P2\_123426/1.
M.\ V.\ M.\ is grateful for support from the National Science Foundation under grant AST-1109247 and from NASA under DPR number NNX11AO41G. 
J.M., M.R. and J.M.P. acknowledge support by the Spanish Ministerio de
Econom\'{\i}a y Competitividad (MINECO) under grants AYA2010-21782-C03-01 and
FPA2010-22056-C06-02. J.M. acknowledges support by MINECO under grant
BES-2008-004564. M.R. acknowledges financial support from MINECO and European
Social Funds through a \emph{Ram\'on y Cajal} fellowship. J.M.P. acknowledges
financial support from ICREA Academia.

Based on observations made with ESO Telescopes at the La Silla Paranal Observatory under programme ID 086.D-0511. This work was supported by the Centre National d'Etudes Spatiales (CNES), based on observations obtained with MINE -the Multi-wavelength INTEGRAL NEtwork.

The authors thank the International Space Science Institute (ISSI, Bern) for
support within the ISSI team ``Study of Gamma-ray Loud Binary Systems'' and
SFI/HEA Irish Centre for High-End Computing (ICHEC) for the provision of
computational facilities and support.





\def\aj{AJ}%
\def\actaa{Acta Astron.}%
\def\araa{ARA\&A}%
\def\apj{ApJ}%
\def\apjl{ApJ}%
\def\apjs{ApJS}%
\def\ao{Appl.~Opt.}%
\def\apss{Ap\&SS}%
\def\aap{A\&A}%
\def\aapr{A\&A~Rev.}%
\def\aaps{A\&AS}%
\def\azh{AZh}%
\def\baas{BAAS}%
\def\bac{Bull. astr. Inst. Czechosl.}%
\def\caa{Chinese Astron. Astrophys.}%
\def\cjaa{Chinese J. Astron. Astrophys.}%
\def\icarus{Icarus}%
\def\jcap{J. Cosmology Astropart. Phys.}%
\def\jrasc{JRASC}%
\def\mnras{MNRAS}%
\def\memras{MmRAS}%
\def\na{New A}%
\def\nar{New A Rev.}%
\def\pasa{PASA}%
\def\pra{Phys.~Rev.~A}%
\def\prb{Phys.~Rev.~B}%
\def\prc{Phys.~Rev.~C}%
\def\prd{Phys.~Rev.~D}%
\def\pre{Phys.~Rev.~E}%
\def\prl{Phys.~Rev.~Lett.}%
\def\pasp{PASP}%
\def\pasj{PASJ}%
\def\qjras{QJRAS}%
\def\rmxaa{Rev. Mexicana Astron. Astrofis.}%
\def\skytel{S\&T}%
\def\solphys{Sol.~Phys.}%
\def\sovast{Soviet~Ast.}%
\def\ssr{Space~Sci.~Rev.}%
\def\zap{ZAp}%
\def\nat{Nature}%
\def\iaucirc{IAU~Circ.}%
\def\aplett{Astrophys.~Lett.}%
\def\apspr{Astrophys.~Space~Phys.~Res.}%
\def\bain{Bull.~Astron.~Inst.~Netherlands}%
\def\fcp{Fund.~Cosmic~Phys.}%
\def\gca{Geochim.~Cosmochim.~Acta}%
\def\grl{Geophys.~Res.~Lett.}%
\def\jcp{J.~Chem.~Phys.}%
\def\jgr{J.~Geophys.~Res.}%
\def\jqsrt{J.~Quant.~Spec.~Radiat.~Transf.}%
\def\memsai{Mem.~Soc.~Astron.~Italiana}%
\def\nphysa{Nucl.~Phys.~A}%
\def\physrep{Phys.~Rep.}%
\def\physscr{Phys.~Scr}%
\def\planss{Planet.~Space~Sci.}%
\def\procspie{Proc.~SPIE}%
\let\astap=\aap
\let\apjlett=\apjl
\let\apjsupp=\apjs
\let\applopt=\ao
\bibliographystyle{mn2e}
\bibliography{Pulsar_Catalog_ALL_Refs_new}

\label{lastpage}
\end{document}